\documentclass[11pt]{article}
\usepackage{moriond,epsfig}

\def\Journal#1#2#3#4{{#1} {\bf #2}, #3 (#4)}

\def\NPB{{\em Nucl. Phys.} B}
\def\PLB{{\em Phys. Lett.}  B}

\def\PRD{{\em Phys. Rev.} D}

\def\be{\begin{equation}}
\def\ee{\end{equation}}
\def\bea{\begin{eqnarray}}
\def\eea{\end{eqnarray}}

\begin{document}
\vspace*{4cm}
\title{New Results in Soft Gluon Physics}

\author{ C.D. WHITE }

\address{School of Physics and Astronomy, University of Glasgow, Glasgow G12 8QQ, Scotland, UK}

\maketitle\abstracts{
We examine soft gluon physics, focusing on recently developed path integral
methods. Two example applications of this technique are presented, namely
the classification of soft gluon amplitudes beyond the eikonal approximation,
and the structure of multiparton webs. The latter reveal new mathematical 
structures in the exponents of scattering amplitudes.}

\section{Introduction}
It is well-known that QCD radiation leads to unstable results in perturbation
theory when the momentum of the emitted radiation becomes low (``soft''). 
Typically, if $\xi$ is some dimensionless energy variable representing the
total energy carried by soft gluons, then one finds differential cross-sections
of the form
    \begin{equation}
      \frac{d\sigma}{d\xi}=\sum_{n,m}\alpha_S^n\left[c^0_{nm}\frac{\log^m(\xi)}{\xi}+c^{1}_{nm}\log^m(\xi)+\ldots\right]
    \label{xsec}
    \end{equation}
involving large logarithms (of soft origin) to all orders in perturbation
 theory.
Here the first set of terms can be obtained from the so-called {\it eikonal 
approximation}, in which the momentum of the emitted gluons formally goes to 
zero. Much is known already about these logarithms. The second set of terms
arises from the {\it next-to-eikonal (NE) limit}, corresponding to a first
order expansion in the momentum of the emitted gluons. These logarithms, 
although suppressed by a power of the energy scale $\xi$, can be numerically
significant in many scattering processes. 

\begin{figure}
\begin{center}
\psfig{figure=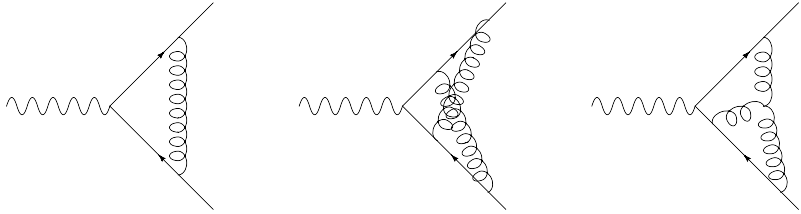,height=1.1in}                                        
\caption{Schematic hard scattering process dressed by soft gluons.
\label{fig:scatter}}
\end{center}
\end{figure}

In the soft phase-space region in which $\xi\rightarrow0$, 
the above perturbation
expansion breaks down in that all terms become large. The solution to this 
problem is to work out the logarithms to all orders in the coupling constant
and sum them up (``resummation''). This is by now a highly developed subject,
and many different approaches already exist for summing eikonal logarithms
(e.g. Feynman diagram approaches, SCET). Here I will explain the
basic idea using the {\it web} 
approach~\cite{Gatheral,FT,Sterman}, and using the schematic scattering
process shown in figure~\ref{fig:scatter}. This consists of a hard interaction
(in this case a virtual photon and quark pair of nonzero momentum), which
is dressed by gluons (we do not distinguish real and virtual emissions
in the figure). When these gluons become soft, this generates the large 
logs in eq.~(\ref{xsec}). However, one may show that the soft gluon
diagrams {\it exponentiate}. That is, if ${\cal A}$ is the amplitude for
the Born interaction ${\cal A}_0$ dressed by any number of soft gluons, one
has~\cite{Gatheral,FT,Sterman}
\begin{equation}
{\cal A}={\cal A}_0\exp\left[\sum \tilde{C}_W W\right],
\label{Aexp}
\end{equation}
where the sum in the exponent is over soft-gluon diagrams $W$. 
This is a powerful result for two reasons.
Firstly, large logs coming from the soft gluon diagrams sit in an exponent,
thus get summed up to all orders in perturbation theory. Secondly, not all 
soft gluon diagrams have to be calculated. It turns out that only those
which are irreducible (``webs'') need to be considered, and the first few
examples are shown in figure~\ref{fig:scatter}. The webs have modified
colour factors $\tilde{C}(W)$, which are not the usual colour factors of
perturbation theory, and these are zero for non-webs. 
We see that crucial to resummation is the notion of
exponentiation, and indeed this is common to all other approaches.

\begin{figure}
\begin{center}
\psfig{figure=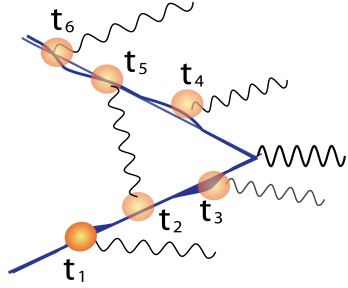,height=1.5in}
\caption{Spacetime depiction of Drell-Yan production, showing the
trajectories of the incoming quarks.
\label{fig:path}}
\end{center}
\end{figure}

Having very briefly reviewed soft gluon physics, let us now focus on the
following open problems:
\begin{enumerate}
\item {\it Can we systematically classify next-to-eikonal logarithms?}
As remarked above, much less is known about the second set of logs in 
eq.~(\ref{xsec}) than the first set. 
A number of groups have looked at this in recent 
years~\cite{LMS,MV,SMVV,GR,LSW,LMSW}. 
\item {\it What is the equivalent of webs for multiparton processes?}
The webs of~\cite{Gatheral,FT,Sterman} are only set up for cases in which
two coloured particles interact e.g. Drell-Yan production, deep inelastic
scattering, $e^+e^-\rightarrow q\bar{q}$ etc. Recent work has tried to
generalise the web concept to processes with many coloured 
particles~\cite{GLSW,MSS}, 
which are ubiquitous at hadron colliders. 
\end{enumerate}
Both of these questions are conveniently addressed using the path integral
technique for soft gluon resummation. 
The essential idea of this approach is that QCD scattering processes are 
rewritten in terms of (first-quantised) path integrals over the trajectories 
of the hard emitting particles. To see what this means in more detail, 
consider the cartoon shown in figure~\ref{fig:path}, which shows Drell-Yan
production, in which incoming quarks fuse to make a final state vector boson.
If we think about this process in position space, the incoming particles
can emit gluons at various places along their spacetime trajectories. We
have already seen that the eikonal approximation (which gives the first 
set of logs in eq.~(\ref{xsec})) corresponds to the emitted gluons having zero
momentum. Then the incoming particles do not recoil, and so follow 
classical straight line trajectories. 
Beyond the eikonal approximation, each trajectory
will get a small kick or wobble upon emission of a gluon. The sum over all
possible wobbles that each trajectory can have is equivalent, in a well-defined
sense, to a sum over possible gluon emissions of nonzero momentum. The sum
over wobbles of a trajectory is nothing other than a Feynman path integral,
as used in Feynman's original formulation of quantum mechanics. Thus, it 
follows that there should be a description of soft gluon physics in terms of
path integrals for the hard external particles, where the leading term of
each path integral (the classical trajectory) gives the eikonal approximation.
If one can then somehow systematically expand about the classical trajectory 
and keep the ``first-order set of wobbles'', this gives the next-to-eikonal 
corrections~\cite{LSW}. With this approach, we have proved that
the structure of NE corrections to scattering amplitudes has the generic form
\begin{equation}
{\cal A}={\cal A}_0\exp\left[{\cal M}^E+{\cal M}^{NE}\right]\times
\left[1+{\cal M}_{rem.}\right]+{\cal O}(NNE).
\label{Aexp2}
\end{equation}
Here the left-hand side denotes the amplitude for a given Born interaction
${\cal A}_0$ dressed by soft and next-to-soft gluons. The first term in the
exponent denotes eikonal webs~\cite{Gatheral,FT,Sterman}, 
and the second term constitutes
next-to-eikonal webs. Finally there is a remainder term whose interpretation
is also understood~\cite{LSW}. The above formula has been confirmed using
an explicit diagrammatic proof, and preliminary calculations
in Drell-Yan production have been carried out which pave the way for
resummation of next-to-eikonal effects~\cite{LMSW}. 
Interestingly, the same schematic structure of 
next-to-eikonal corrections also holds in perturbative quantum 
gravity~\cite{Gravity}.
\begin{figure}
\begin{center}
\psfig{figure=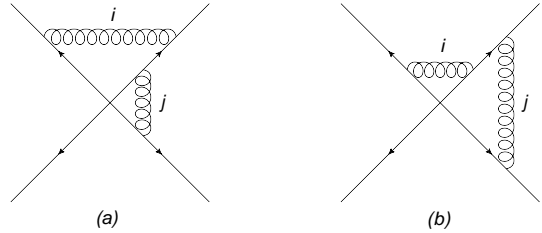,height=1.2in}
\caption{Example web at two loop order, for a four parton process.
\label{fig:web}}
\end{center}
\end{figure}

We now turn to the second open problem above, that of 
generalising webs (diagrams
which sit in the exponent of the soft gluon amplitude) from 
two parton to multiparton scattering. In the two parton case of 
eq.~(\ref{Aexp}), we saw that webs were single (irreducible) diagrams.
This becomes more complicated in multiparton processes: webs are no longer
irreducible, but become compound sets of diagrams, related in 
a particular way. Consider, for example, the two diagrams shown in 
figure~\ref{fig:web}, which are a two-loop soft gluon correction 
to a hard interaction
involving four partons. Taking diagram (a), we can make a second
diagram by permuting the gluons on the upper right-hand line. This gives
diagram (b), and by performing the permutation again we get back the
original diagram. The graphs thus form a closed set under 
permutations of gluon emissions. Such closed sets are argued to be the
appropriate generalisation of webs to multiparton scattering~\cite{GLSW}. 
The derivation of these results uses the {\it replica trick}, an elegant 
technique for proving exponentiation properties which is borrowed from
statistical physics.

Each diagram $D$ in a given closed set (web) has a kinematic part ${\cal F}(D)$
and a colour factor $C(D)$. In the normal amplitude, these are simply 
multiplied together. However, in the exponent of the amplitude, the colour
and kinematic parts of web diagrams mix with each other. That is, a single
web contributes a term
\begin{equation}
  \sum_{D,D'}{\cal F}_D R_{DD'}C_{D'}
\end{equation}
to the soft gluon exponent, where the sum is over diagrams in the web,
and $R_{DD'}$ is a {\it web mixing matrix} which describes how the vectors
of kinematic and colour factors are entangled. The study of multiparton webs
is thus entirely equivalent to the study of web mixing matrices. They are 
matrices of constant numbers (e.g. independent of the number of colours) 
that encode a huge amount of physics! An ongoing goal is to classify general
properties of these matrices, and to translate these into physical results.

We already know about some interesting properties. Firstly, any row
of any web mixing matrix has elements which sum to zero. Secondly, any
web mixing matrix is idempotent, that is $R^2=R$. The
matrices are thus projection
operators, having eigenvalues of 0 and 1 (with an appropriate 
degeneracy). These
properties have been interpreted physically~\cite{GLSW}, and the proofs
use both the replica trick and known properties of combinatorics~\cite{GW}. 
This latter point is itself interesting,
as a pure mathematician could have proved these results without in fact knowing
any of the underlying physics. This suggests that there are two ways of
finding out more about web mixing matrices - either one may apply known
physics constraints and see what this implies in web mixing matrix language,
or one may study the matrices from a pure combinatorics point of view, and 
learn in the process about the entanglement of colour and kinematics
\footnote{Note that extra complications arise when the renormalisation of 
webs is considered~\cite{MSS}.}!

To summarise, path integral methods prove highly powerful in analysing
soft gluon physics, allowing new results to be obtained. Specifically,
we have outlined the classification of next-to-eikonal corrections, and also
the structure of multiparton webs. The results have application to the
resummation of logarithms in cross-sections, but may also have more formal
applications in elucidating the structure of scattering amplitudes in a 
variety of field theory contexts. Investigation of these possibilities is
ongoing.

\section*{Acknowledgments}
I wish to thank the organisers of the Moriond QCD session for a very 
enjoyable conference, and am also grateful to my collaborators 
Einan Gardi, Eric Laenen, Lorenzo Magnea 
and Gerben Stavenga. This research was supported by the STFC Postdoctoral
Fellowship ``Collider Physics at the LHC''.

\end{document}